 \documentstyle[epsfig]{aipproc}
 
\begin{document}
 
 \title{X-ray Observations of Young Supernova Remnants}
 
 \author{Una Hwang}
 \address{NASA Goddard Space Flight Center, Greenbelt, MD 20771, and\\ University of Maryland, College Park, MD 20742, USA}

 \maketitle
 
 \begin{abstract}
 
 This brief review of recent X-ray observations of young supernova
 remnants highlights results obtained by the Chandra and XMM Newton
 Observatories since their launch last year. Their impressive
 capabilities are illustrated by results for spectral imaging, and for
 spatially resolved spectroscopy that isolates emission from individual
 ejecta knots and from the \index{shock, blast} forward shock. I also review X-ray
 dynamical studies of supernova remnants, which should undergo
 significant advances during this new era.
 
 \end{abstract}
 \hyphenation{study}
 \hyphenation{ele-ment}
 
 \section*{Introduction}
 
 Young supernova remnants are expected to emit brightly in X-rays, as
 the forward and \index{shock, reverse} \index{shock, reverse} reverse shocks in remnants typically heat the ambient
 gas and ejecta to X-ray emitting temperatures. Moreover, supernova
 remnants are expected to be rich X-ray emission line sources: the gas
 ionizes slowly after being suddenly heated behind the shock, so that
 the populations of certain stable ions can be significantly higher
 than would be expected at equilibrium for a given electron
 temperature. The result is correspondingly enhanced line emission
 from those ions. For remnants whose X-ray emission is dominated by
 the \index{shock, reverse} reverse-shocked ejecta, the line emission will be further enhanced
 by the high element abundances in the ejecta that result from nuclear
 processing, both before and during the supernova explosion.
 
 It has been twenty years since early X-ray spectral observations at
 both moderate and high resolution first verified the rich X-ray
 emission line spectra of supernova remnants (e.g., Becker et
 al. \cite{becker}, Winkler et al. \cite{winkler}); meanwhile, ever
 improving X-ray images have revealed their beautiful and complex
 morphologies (e.g., see Seward \cite{seward} and references therein,
 Aschenbach et al. \cite{aschenbach}, Levenson et
 al. \cite{levenson97}, Williams et al. \cite{williams99}).
 Opportunities for simultaneous imaging and spectroscopy, however, have
 been limited until relatively recently. The launch of the ASCA
 satellite in 1993 \cite{tanaka} marked a significant advance, in that
 $1'$ optics were coupled with CCD imaging spectrometers of moderate
 spectral resolution that were capable of isolating certain emission
 lines and blends. ASCA thus provided the first X-ray spectral images
 \cite{holt,fuji}, as well as important spatially resolved spectral
 results for remnants of larger angular size.
 
 In 1999, the launch of Chandra and XMM Newton provided an enormous
 further advance in that the CCD spectrometers were placed at the focus
 of far better mirrors---XMM Newton provides a point source image with
 a roughly $6''$ core, while Chandra provides a $0.5''$ angular
 resolution that is the best ever realized for X-rays.
 The two missions are complementary, with Chandra's strength
 being its superior mirrors, and XMM's strength being its higher
 collecting area, particularly at energies above 5 keV (for example,
 see the impressive very high energy X-ray image of Cas A taken by XMM
 \cite{bleeker}). Both missions also include dispersive high
 resolution spectrometers, with complementary characteristics, that
 have already provided exciting results for supernova remnants in the
 Magellanic Clouds (see the review by Canizares \cite{canizares2000}).
 Unfortunately, such observations are difficult for remnants of any
 significant angular size because of the complicated blending of highly
 detailed spatial and spectral information.
 
 This review highlights a few of the many results obtained by Chandra
 and XMM during their first year, with the goal of illustrating the
 power of simultaneous imaging and spectroscopy with moderate
 resolution CCD spectrometers. It is restricted to a few illustrative
 examples of spectral imaging and of detailed spectroscopy of compact
 regions of thermal X-ray emission from shocked ejecta and ambient gas.
 Numerous other such examples have been presented at this conference.
 I conclude by looking forward to the coming of age of X-ray dynamical
 studies of supernova remnants.
 
 \section*{X-ray Spectral Imaging}
 
 The capability to image a source in specific spectral features from a
 single element is particularly important for young supernova remnants
 whose emission is dominated by their ejecta, as such images can shed
 light on the different physical conditions and spatial distribution of
 various elements in the ejecta. Use of such techniques has been
 commonplace at optical wavelengths, but optically emitting ejecta
 generally represent a relatively small fraction of the total shocked
 ejecta mass; few remnants even have significant optical emission from
 their ejecta to begin with, excepting Cas A and other so-called
 ``oxygen-rich'' remnants. The wide bandpass of the ASCA, Chandra, and
 XMM CCD spectrometers also allows the comparison of relatively hard
 (4--6 keV) continuum images with softer ($\lesssim$ 2 keV) line
 images, which has been useful in identifying emission from the forward
 shock, as discussed below.
 
 Later in this paper, we consider spectral imaging results from Chandra
 in conjunction with spectroscopic results. Here we consider the
 recent XMM Newton observation of Tycho's SNR.
 
 \subsection*{Tycho's SNR with XMM}
 
 Tycho's supernova remnant (SN 1572), with its $8'$ diameter on the
 sky, is ideally suited for XMM, with its 37$'$ field of view and 6$''$
 PSF core. Tycho's X-ray spectrum is notable for the richness of its
 line emission, from Fe L transitions (to the n=2 level) with energies
 near 1 keV, and an exceptionally prominent Si He $\alpha$ blend near
 1.86 keV, to Fe K transitions (to the n=1 level) with energies near
 6.4 keV. An imaging study with the ASCA satellite by Hwang \&
 Gotthelf \cite{hg} suggested that the Fe L and Fe K emission are not
 spatially coincident, with the Fe K emission at a smaller average
 radius than other emission lines. That earlier study is limited by
 both photon statistics and angular resolution, but new XMM images
 presented by Decourchelle et al. \cite{decourchelle} clearly show that
 the azimuthally averaged radial distribution of Fe K emission has a
 peak that lies well inside that of the Fe L emission. By contrast,
 the Fe L and Si He $\alpha$ emission are consistent with having the
 same average radial distribution.
 
 
 These differences in the radial profiles may be qualitatively
 understood as arising from the temperature and ionization structure of
 the ejecta behind the \index{shock, reverse} reverse shock. The ejecta of Type \index{supernovae, Type Ia } Ia supernovae
 have a steep outer density profile surrounding an inner plateau of
 more constant density. As the \index{shock, reverse} reverse shock reaches the inner plateau
 in the ejecta distribution, the temperature of the shocked gas should
 increase and the density decrease. The more recently reverse-shocked
 gas is therefore hotter and has a lower ionization age, as would be
 required to explain the presence of Fe K emission at smaller radii, as
 pointed out by Decourchelle et al. \cite{decourchelle}. The low
 density, hot Fe emission has also been suggested to be associated with
 \index{radioactive heating} radioactively heated \index{bubbles, iron-nickel } bubbles of Fe ejecta by Wang \& Chevalier
 \cite{wangchevalier}. Detailed further analysis of observations with
 Chandra and XMM should shed further light on these issues.
 
 \section*{Spatially Resolved Spectroscopy}
 
 For young remnants whose emission is dominated by their
 \index{shock, reverse} reverse-shocked ejecta, the capability to do detailed spatially
 resolved X-ray spectroscopy with Chandra and XMM is a wonderful
 opportunity to study the X-ray emitting ejecta. The XMM observation
 of Tycho's SNR, for example, straighforwardly confirms the very
 different compositions of Si and Fe in the two bright knots at the
 eastern periphery of the remnant \cite{decourchelle}, a result
 indicated previously by Vancura et al. \cite{vancura}, but only with
 considerable data manipulation and luck using the limited instruments
 then available. XMM isolates the spectra of these knots for the first
 time. The superior angular resolution of the Chandra mirrors allows
 one to probe the spectra of even more compact regions, and for this
 illustration we turn to the brightest Galactic remnant, Cas A.
 
 \subsection*{Ejecta Knots in Cas A}
 
 Cas A has long been considered the undisputed product of the explosion
 of a massive star, probably a \index{Wolf-Rayet stars} Wolf-Rayet star
 \cite{fesen,fesengunderson}, though the compact stellar remnant that
 then must have been produced remained elusive until Chandra's first
 light revealed it in X-rays \cite{tananbaum,pavlov,chakrabarty}. This
 brief initial observation with the Advanced CCD Imaging Spectrometer
 (ACIS) also showed that the individual ejecta knots in Cas A have a
 range of spectral characteristics (Hughes et al. \cite{hughescasa}),
 some of which are qualitatively consistent with their origin in
 different explosive \index{nucleosynthesis } nucleosynthesis burning zones of a massive star.
 For instance, the spectrum of a typical knot with emission lines of
 Si, S, Ar, and Ca was shown to be consistent with the element
 abundances expected for the zone of explosive O burning, whereas
 another knot also showing Fe emission is consistent with the zone of
 \index{explosive burning, of Si} incomplete Si burning. A region in the outer southeast region of the
 remnant that was examined by Hughes et al. required the presence of
 still more Fe in the ejecta, suggesting a macroscopic mixing and
 inversion of the Si rich ejecta layer and the innermost layer of Fe.
 
 An observation of Cas A twenty times deeper than the first light
 observation was also obtained by Chandra. With this observation, the
 distributions of the Si, S, Ar, Ca, and Fe-emitting ejecta were mapped
 through the equivalent widths of their emission lines by Hwang, Holt,
 \& Petre \cite{hwang}. The location of Fe-emitting ejecta mapped in
 this way is exterior to that of the Si-emitting ejecta throughout the
 southeast region, but not elsewhere. Because of this singularity of
 the southeast region, we turn to a more detailed spectral analysis 
 using the deep Chandra observation.
 
 Figure 1 shows the southeastern part of Cas A and indicates the
 regions studied; the spectral extraction regions were as small as
 $3''$ in diameter. The spectra were analyzed using models for plane
 parallel shocks in XSPEC version 11.0 \cite{borkowskisedov} with a
 single electron temperature, and including a range of ionization ages
 from zero up to a fitted maximum value. The ionization age $n_et$ is
 defined as the product of the ambient electron density and the time
 since the gas was heated by the shock, with equilibrium corresponding
 to roughly $10^{12}$ cm$^{-3}$ s or higher. Also indicated in the
 figure are the fitted Fe/Si abundance ratios for the various knots,
 classified broadly as dominated by Si (Fe/Si $< 0.1$
 Fe$_\odot$/Si$_\odot$), dominated by Fe (Fe/Si $> 1.5$
 Fe$_\odot$/Si$_\odot$) and with a mixture of Fe and Si
 (0.1 Fe$_\odot$/Si$_\odot <$ Fe/Si $<$ 1.5 Fe$_\odot$/Si$_\odot$). The
 highest Fe/Si abundance ratios systematically occur in the outermost
 regions in the southeast, with fitted ratios as high as
 $4-5$ Fe$_\odot$/Si$_\odot$ in the regions that we examined. An
 intermediate region with a more solar mixture of Si and Fe gives way
 to a Si-dominated region coincident with the bright ring evident in
 the broadband images.
 
 Two extreme examples of the spectra for a {\it single} highly Fe-rich
 knot and a more typical S-rich knot are shown in Figure 2. Both have
 a fitted electron temperature of kT $\sim$ 2 keV, that is typical of
 the ejecta knots. What is obviously very different between the two is
 the relative abundance of Fe to Si (roughly a factor of 200 higher in
 the spectrum on the left), that manifests itself in the shape of the
 continuum below the Si blends. Because of the high Fe abundance in
 the spectrum on the left, this continuum comes predominantly from Fe
 rather than from H and He. The ionization age $n_et$ is also some 50
 times higher for the Fe-rich knot than it is for the Si-rich knot.
 The high ionization age is evident in the shape of the Fe L emission
 in the spectrum on the left, and is also reflected in the increased
 strength of the Si Ly $\alpha$ (2.006 keV) line relative to the Si He
 $\alpha$ ($\sim$ 1.85 keV) blend. Assuming that, to zeroth order, the
 Fe and Si knots were shocked at roughly the same time, this indicates
 much higher densities in the Fe-rich knots, consistent with their
 origin deep within the inner layers of the supernova ejecta.
 
 \begin{figure} 
 \centerline{\epsfig{file=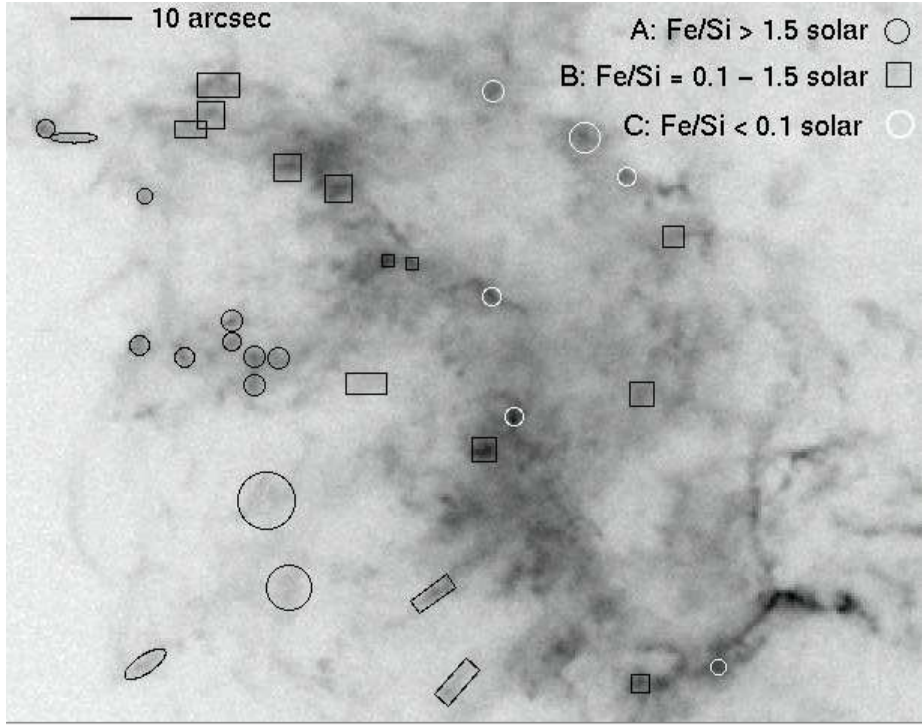,height=3.8in}}
 \vspace{10pt}
 \caption{Closeup view of the southeastern section of Cas A with
 Chandra ACIS in a deep 50 ks observation (region is $2.5'$ across).
 Spectral extraction regions are overlaid and coded according to the
 fitted Fe/Si element abundance ratio (relative to solar): A (high Fe)
 circular and elliptical regions, B (mixture) square and
 rectangular regions, C (high Si) white circular regions.}
 \end{figure}
 
 \begin{figure}
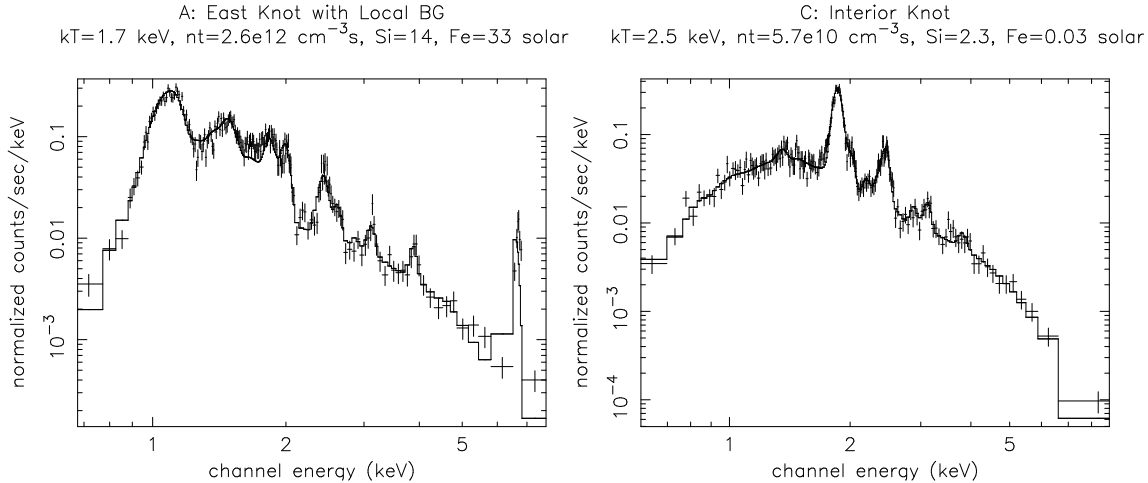
 
 \centerline{\epsfig{file=e2a_a.ps,width=2.5in,angle=-90}\epsfig{file=i3_c.ps,width=2.5in,angle=-90}}
 \vspace{10pt}
 \caption{Representative spectra of extremely Fe- (left) and
 Si-dominated (right) knots in the southeast region of Cas A. The the
 best-fit \index{shock, plane-parallel} plane-parallel shock model with a single electron temperature
 is overlaid on the data, and fitted parameter values are indicated.}
 \end{figure}
 
 The region of the northeast jet, the largest radial extension seen in
 Cas A, is well-known from detailed optical observations to be enriched
 by ejecta emitting in O, S, and Ar \cite{fesen,fesengunderson}. The
 Chandra data show that the X-ray emission arises largely from ejecta
 highly enriched in Si, Si, Ar, and Ca. Fits to their X-ray spectra
 also show higher ionization ages than in the bright ring, suggesting
 that these are also high density clumps flung outward at high velocity
 during the explosion.
 
 \section*{Isolation of Emission from the Forward Shock}
 
 It was evident from the earliest nonimaging spectral observations that
 the \index{shock, reverse} reverse-shocked ejecta in Tycho and Cas A are the dominant
 contributors to the soft X-ray emission at energies below about 5 keV
 \cite{becker80,becker}. The \index{shock, reverse} reverse shock cannot exist, however,
 without the \index{shock, blast} forward shock. For Cas A and Tycho, spatial X-ray
 components corresponding to the \index{shock, blast} forward shock were suggested from
 early Einstein observations, for example by Fabian et
 al. \cite{fabian} and Seward, Harnden \& Tucker \cite{sewardtycho},
 and hard components in the spectra were attributed early on to the
 \index{shock, blast} forward shock (e.g. Pravdo \& Smith \cite{pravdo}. The true situation
 is more complex because theory also predicts nonthermal X-ray emission
 from \index{particle acceleration } particle acceleration at the \index{shock, blast} forward shock and such emission is
 now observed, with the most famous example being SN1006 \cite{koyama}.
 To unambiguously identify the location of the \index{shock, blast} forward shock and to
 isolate its spectrum, one requires the capability to disentangle
 spatial and spectral information that is now available with Chandra
 and XMM.
 
 For Cas A, the Chandra X-ray continuum image at energies between $4-6$
 keV shows a ``rim'' surrounding the bright ring of emission from the
 ejecta, as shown by Gotthelf et al. \cite{gotthelf}. As discussed
 elsewhere in these proceedings by Rudnick \cite{rudnick}, this is
 persuasively identified as the location of the \index{shock, blast} forward shock. The
 location of the \index{shock, blast} forward shock in Tycho is similarly identified by XMM
 from the good correlation of this X-ray continuum emission with radio
 emission \cite{decourchelle}.
 
 At distances of a few kpc within the Galaxy, Cas A and Tycho are
 relatively nearby, but Chandra's spatial resolution is good enough to
 isolate the \index{shock, blast} forward shock in much more distant remnants. The bright
 \index{ejecta-dominated } ejecta-dominated ring of E0102-72, the brightest remnant in the Small
 Magellanic Cloud, has been previously imaged \cite{hughes88} ; Chandra
 reveals a faint shelf of emission exterior to the ejecta ring that is
 identified as the \index{shock, blast} forward shock in the observation by Gaetz et
 al. \cite{gaetz}. The spectrum of this outer region is qualitatively
 different from that of the ring, as demonstrated by Hughes, Rakowski,
 \& Decourchelle \cite{hughese0102}, with line strengths that are
 well-explained by the ambient abundances in the SMC, in contrast to
 the prominent O and Ne emission from the ejecta-enriched ring.
 Importantly, the measured temperature of the blast-shocked gas is
 barely 1 keV. Together with the \index{shock velocity} shock velocity inferred from the
 measured X-ray expansion of the \index{shock, blast} \index{shock, blast} forward shock, this has the
 interesting implication that the \index{cosmic ray acceleration} acceleration of cosmic rays at the
 \index{shock, blast} forward shock is highly efficient, as the shock should have heated the
 electrons to substantially higher temperature, even if the electrons
 equilibrate to the much higher ion temperatures only by Coulomb
 collisions. It would therefore appear that some of the enery of the
 shock wave has been diverted elsewhere, with \index{particle acceleration } particle acceleration
 being a persuasive explanation.
 
 \section*{X-ray Dynamical Studies of Remnants}
 
 A frontier that will be explored productively, especially by Chandra,
 is the study of the dynamics of the X-ray emitting gas in remnants. The
 first X-ray \index{proper motion } proper motion studies of the youngest and brightest
 supernova remnants were carried out using the high resolution imaging
 experiments on the Einstein and ROSAT Observatories, generally for
 broad radial or azimuthal sectors (Kepler's and Tycho's SNRs by Hughes
 \cite{hugheskepler,hughestycho}, Cas A by Vink et al. \cite{vink} and
 Koralesky et al. \cite{koralesky}). The general trends in these
 results are two-fold. First, in comparison to similar radio expansion
 studies \cite{dickel,reynoso,anderson,agueros}, the X-ray expansion
 results have yielded velocities that are systematically higher by
 roughly a factor of two. Second, in spatially resolved expansion
 studies carried out in both the radio and in X-rays, the expansion is
 observed to be significantly asymmetric, with expansion rates varying
 with azimuthal angle by factors of two or even significantly more.
 Both these points are illustrated in Figure 3 of Rudnick's review of
 Cas A elsewhere in these proceedings.
 
 The disparity between the radio and X-ray expansion results is
 disconcerting. In most instances, the extent of the radio and X-ray
 emission from the remnants is similar, and yet the rate of radio
 expansion is only about half the rate of the X-ray expansion. Further
 observations with Chandra over the next several years should help to
 shed light on this puzzle. In particular, dynamical studies for the
 brightest remnants could be carried out for individual X-ray knots,
 and moreover the analysis could be done for each element separately, as
 has now been done extensively (at least for Cas A and a few other
 remnants) at optical and radio wavelengths.
 
 The dynamics in the direction along the line of sight can be studied
 by measuring this velocity component through \index{Doppler shifts } Doppler shifts in the
 energies of known emission lines. For Cas A, the presence of
 asymmetries in the line of sight velocities was known from early
 observations with the Focal Plane Crystal Spectrometer (FPCS) on the
 Einstein Observatory, a dispersive spectrometer that was used in this
 case with a $3' \times 30'$ aperture. Markert et al. \cite{markert}
 show that the FPCS spectral scans of He-like Si and S resonance and
 H-like Ly $\alpha$ lines from the northwest half of Cas A have lower
 line energies than those from the southeast, implying bulk motion of
 the gas in a roughly ring-shaped geometry. A \index{Doppler shifts } Doppler map by Holt et
 al. \cite{holt} based on the centroid of the Si He $\alpha$ blend
 using more recent ASCA data (therefore with lower spectral resolution
 than the FPCS) confirms and extends this result with an angular
 resolution of about $1'$.
 
 The Chandra version of the \index{Doppler shifts } Doppler velocity map of Cas A obtained from
 the centroids of the Si He $\alpha$ blend is shown in Figure 3. The
 energy shifts measured correspond to velocities as high as 2500
 km/s, with \index{Doppler shifts } blue-shifted material generally located in the southeast
 and \index{Doppler shifts } red-shifted material in the northwest, consistent with
 previous measurements. The Chandra map gives significantly more
 detail than previous maps: in the northwest, a concave arc of
 moderately \index{Doppler shifts } blueshifted material is seen just south of a convex arc of
 \index{Doppler shifts } redshifted material. These structures qualitatively correspond very
 well with the velocity structure of the dense ejecta knots measured
 with optical echelle spectra by Lawrence et al. \cite{lawrence},
 though there is more X-ray emission in the southeast than in the optical. A
 range of ionization ages is seen in the Si-emitting material in Cas A,
 and in principle this could affect the centroid of the Si blend, but
 fortunately the Si energy centroid remains dominated by the resonance
 line for the relevant range of ionization age and temperature in Cas
 A. The energy shifts therefore are dominated by \index{Doppler shifts } Doppler shifts due to
 the bulk motion of the gas. Even with moderate energy resolution CCD
 spectrometers, Chandra provides more details of the line-of-sight
 velocity structure in the X-ray emitting gas than was previously
 available.
 
 \begin{figure} 
 \centerline{\epsfig{file=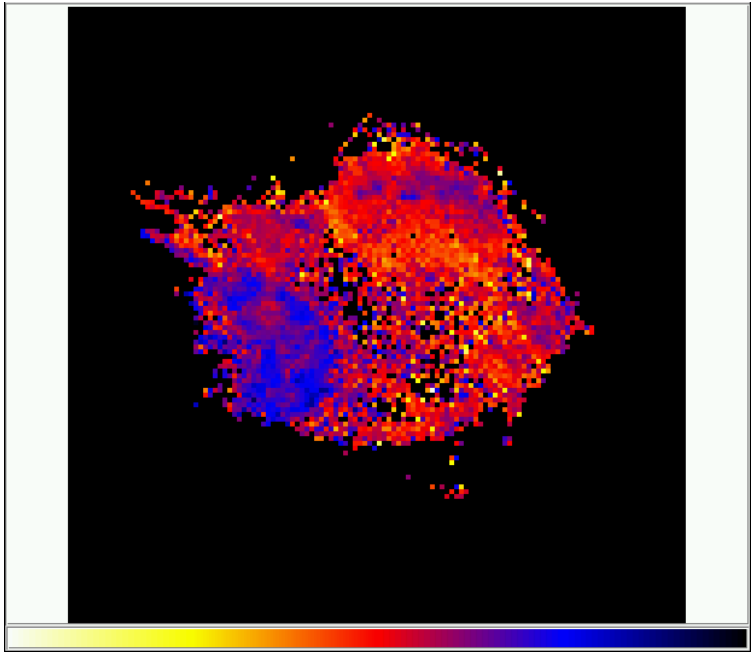,height=3.4in}}
 \vspace{10pt}
 \caption{Doppler velocity map of Cas A based on the Si He $\alpha$
 blend observed with ACIS on Chandra. Black corresponds to blueshifted
 velocities, white to \index{Doppler shifts } redshifted velocities. The range of velocity
 shifts measured is roughly $\pm$2500 km/s.}
 \end{figure}
 
 \section*{Conclusions}
 
 Chandra and XMM Newton have opened the way for true spatially resolved
 X-ray spectroscopy of supernova remnants. Though the CCD
 spectrometers are limited in their spectral resolution, the ability to
 disentangle spatial components that change on very small angular
 scales provides a new window in understanding these complex sources.
 The results presented in this review are just the beginning!
 
 \bigskip
 \noindent {\bf Acknowledgements:} It is a pleasure to acknowledge
 Steve Holt, Rob Petre, and Andy Szymkowiak as my collaborators
 on the Chandra observation of Cas~A, and to thank Anne Decourchelle
 for generously providing the results of her XMM study of Tycho's
 supernova remnant prior to publication. We are grateful to Kazik
 Borkowski for fruitful scientific discussions and for making his
 spectral models publicly available.
 
 
 

 \end{document}